\documentclass{jltp}
\usepackage{bm}
\usepackage{graphics}
\usepackage[dvips]{color}
\usepackage{dcolumn}

\renewcommand{\r}{{\bf r}}
\newcommand{\rb}{{\bf r}_{\bot}}
\renewcommand{\k}{{\bf k}}
\newcommand{\p}{{\bf p}}
\renewcommand{\v}{{\bf v}}
\newcommand{\be}{\begin{eqnarray}}
\newcommand{\ee}{\end{eqnarray}}
\newcommand{\tn}{\tilde{n}}

\title{Hydrodynamics of Superfluid Bose Gases in an Optical Lattice at Finite Temperatures}
\author{Satoru Konabe and Tetsuro Nikuni}
\address{Department of Physics, Faculty of Science, Tokyo University of Science, \\
1-3 Kagurazaka, Shinjuku-ku, Tokyo, Japan, 162-8601}
\runninghead{S. Konabe and T. Nikuni}{Hydrodynamics of Superfluid in an Optical Lattice}

\begin{document}
\maketitle
\begin{abstract}
Starting from an effective action for the order parameter field, we derive a coupled set of generalized hydrodynamic equations for a Bose condensate in an optical lattice at finite temperatures.
Using the linearized hydrodynamic equations, we study the microscopic mechanism of the Landau instability due to the collisional damping process between condensate and noncondensate atoms.
It is shown that the Landau criterion of the superfluidity for the uniform system is modified due to the presence of the periodic optical lattice potential.
\end{abstract}

\section{Introduction}
Recently Bose condensed gases in an optical lattice are extensively studied both experimentally and theoretically.  
Much of attention is attracted to the breakdown of the superfluidity in a condensate. 
A lot of works have discussed the stability condition of the superfluidity by using the zero-temperature Gross-Pitaevskii equation.~\cite{wu,machholm2003,kramer2003,menotti2003} 
As emphasizing in our previous works,~\cite{konabe2006_2} it is crucial to explicitly include the degree of freedom of noncondensate atoms into the theory to understand the Landau instability from the microscopic point of view.

In the present paper, by means of an effective action and an associated a coupled set of generalized hydrodynamic equations, we study the microscopic mechanism of the breakdown of the superfluidity having experiments such as reported in Ref.~\onlinecite{sarlo2005} in mind. 

\section{Generalized Hydrodynamic Equations in an Optical Lattice}\label{sec:generalized_hyd}
We consider the Bose condensate at finite temperatures trapped in a one-dimensional optical lattice in addition to the harmonic potential
\be
V_{\rm ext}(\r)=\frac{m}{2}(\omega_x^2x^2+\omega_y^2y^2+\omega_z^2z^2)+sE_{R}\cos^2\left(\frac{\pi}{d}z\right),\label{generalized:potential}
\ee
where $\omega_x$, $\omega_y$, $\omega_z$ are the frequencies of the harmonic trap potential, $s$ is the strength of the lattice potential, $E_{R}$ is the recoil energy, and $d$ is a lattice constant.
In the external potential represented by Eq.~(\ref{generalized:potential}), the Bose condensates are described by the following action
\be
S&=&\int d\r\int dt\ \phi^*(\r,t)\biggl[
i\hbar\frac{\partial}{\partial t}+\frac{\hbar^2}{2m}\nabla^2-V_{{\rm ext}}(\r)
-\frac{g}{2}|\phi(\r,t)|^2\nonumber\\
&&{}-2g\tn(\r,t)\biggl]\phi(\r,t)\nonumber\\
&&{}+\frac{g^2}{2}\int d\r d\r'\int_{c} dtdt'\phi^*(\r,t)F(\r,t;\r',t')\phi(\r',t'),\label{dissipative:action}
\ee
where $\phi(\r,t)$ is a complex field that represents the degree of freedom of condensate atoms, $\tn(\r,t)\equiv\langle\tilde\psi^{\dag}(\r,t)\tilde\psi(\r,t)\rangle$ is the noncondensate density with $\tilde\psi(\r,t)$ being the noncondensate field operator, and $F(\r,t;\r',t')\equiv G(\r,t;\r',t')G(\r,t;\r',t')G(\r',t';\r,t)$ is defined in terms of nonequilibrium Green's functions of the noncondensate atoms 
$
G(\r,t;\r',t')\equiv-i\langle \tilde\psi(\r,t)\tilde\psi^{\dag}(\r',t')\rangle.
$
In Eq.~(\ref{dissipative:action}), $g$ is a coupling constant which is related to the $s$-wave scattering length of atoms. 
We note that the index $c$ of integral in Eq.~(\ref{dissipative:action}) means that time integrations should be performed on Keldysh contour path.~\cite{danielewicz1984}
The derivation and detailed discussion of the action~(\ref{dissipative:action}) will be presented elsewhere.~\cite{konabe2006_3} 

Considering a relatively deep lattice potential with large $s$, we introduce the ansatz (tight-bindeing approximation)
\be
\phi(\r,t)=\sum_{l}w_l(z)\phi_l(\r_{\bot},t)e^{i\theta_l(\r_{\bot},t)},\label{generalized:ansatz}
\ee
for the order parameter field, where $\theta_l(\rb,t)$ is the phase of the order parameter at the $l$th lattice cite, and $w_l$ and $\phi_l$ are real functions. Here we assume the periodicity for $w_l(z)=w_0(z-ld)$, where $w_0$ is localized near $z=0$.
Substituting the ansatz (\ref{generalized:ansatz}) into the action~(\ref{dissipative:action}) and performing a smoothing procedure from discrete variables to continuous ones following Ref.\onlinecite{kramer2002}, we obtain after some algebra 
\be
S&=&\int dt\int d\r\biggl[
\sqrt{n_c(\r,t)}i\hbar\frac{\partial}{\partial t}\sqrt{n_c(\r,t)}-\hbar n_c(\r,t)\frac{\partial \theta(\r,t)}{\partial t}\biggl]\nonumber\\
&&{}
+J\int dt\int d\r n_c(\r,t)\cos\left(d \frac{\partial \theta(\r,t)}{\partial z}\right)\nonumber\\
&&{}+\frac{\hbar^2}{2m}\int dt\int d\r\sqrt{n_c(\r,t)}\biggl\{
\nabla_{\bot}^2\sqrt{n_c(\r,t)}-\sqrt{n_c(\r,t)}\left(\nabla_{\bot}\theta(\r,t)\right)^2\nonumber\\
&&{}+i\biggl[
2\nabla_{\bot}\sqrt{n_c(\r,t)}\cdot\nabla_{\bot}\theta(\r,t)+\sqrt{n_c(\r,t)}\nabla_{\bot}^2\theta(\r,t)\biggl]\biggl\}\nonumber\\
&&{}-\int dt\int d\r\ V_{\rm ho}(\r)n_c(\r,t)
-\frac{\tilde g}{2}\int dt\int d\r\ n_c^2(\r,t)\nonumber\\
&&{}
-\tilde J\int dt\int d\r\  n_c^2(\r,t)\cos\left(d\frac{\partial\theta(\r,t)}{\partial z}\right)\nonumber\\
&&{}-2\tilde g\int dt\int d\r\ n_c(\r,t)\tn(\r,t)\nonumber\\
&&{}+\frac{\tilde g^2}{2}\int dtdt'\int d\r d\r \sqrt{n_c(\r,t)}e^{-i\theta(\r,t)}F(\r,t;\r',t')\sqrt{n_c(\r',t')}e^{i\theta(\r,t)},\nonumber\\
\label{derivation_smoothed:smoothed_action}
\ee
where $\tilde g\equiv gd\int dz w_0^4(z)$ is the effective renormalized coupling constant, $V_{\rm ho}(\r)$ is the harmonic potential in Eq.~(\ref{generalized:potential}), and
$n_c(\r,t)=\phi_l^2(\rb,t)/d$ and $\theta(\r,t)=\theta_l(\rb,t)$ represent the density and phase after the smoothing procedure.
In Eq.~(\ref{derivation_smoothed:smoothed_action}), we defined the following quantities:
\be
J&\equiv&\int dz\ w_l(z)\biggl[\frac{\hbar^2}{2m}\frac{\partial^2}{\partial z^2}-V_{\rm opt}(z)\biggl]w_{l+1}(z),\label{derivation:tunneling_rate}\\
\tilde J&\equiv& 4gd\int dz w_l^3(z)w_{l+1}(z),
\ee
where $V_{\rm opt}(z)$ is the optical lattice potential in Eq.~(\ref{generalized:potential}).

The hydrodynamic equations are derived by imposing the stationary condition on the smoothed effective action~(\ref{derivation_smoothed:smoothed_action}) with respect to arbitrary variations of the density $n_c(\r,t)$ and of the phase $\theta(\r,t)$.
After performing the Markovian approximation and the gradient expansion for the non-local term related to $F(\r,t;\r',t')$, we obtain
\be
&&m\frac{\partial\v_c(\r,t)}{\partial t}+\nabla\biggl[
\frac{m}{2}\frac{m}{m^*_{\mu}}v_z^2(\r,t)+\frac{m}{2}\v_{\bot}^2(\r,t)+V_{\rm ho}(\r)\nonumber\\
&&{}+\tilde g n_c(\r,t)+2\tilde g \tn(\r,t)\biggl]=0,\label{derivation_dissipative:Josephson_eq}\\
&&\frac{\partial n_c(\r,t)}{\partial t}+\frac{\partial}{\partial z}\left[
\frac{m}{m^*_{\mu}}n_c(\r,t)v_z(\r,t)\right]
+\nabla_{\bot}\left(n_c(\r,t)\v_{\bot}(\r,t)\right)=-\Gamma_{12}(\r,t).\label{derivation_dissipative:continuity_eq}\nonumber\\
\ee
where $\Gamma_{12}(\r,t)\equiv 2n_c(\r,t)R(\r,t)$ with
\be
R(\r,t)&\equiv&
\frac{\pi \tilde g^2}{2\hbar}\int\frac{d\k_1}{(2\pi)^3}\frac{d\k_2}{(2\pi)^3}\frac{d\k_3}{(2\pi)^3}\nonumber\\
&&{}\times\delta(\omega_c+\tilde\omega_1-\tilde\omega_2-\tilde\omega_3)
\delta(\k_c+\k_1-\k_2-\k_3)
\nonumber\\
&&{}\times\biggl\{[1+f(\k_1,\r,t)][1+f(\k_2,\r,t)]f(\k_3,\r,t)\nonumber\\
&&{}\quad
-f(\k_1,\r,t)f(\k_2,\r,t)[1+f(\k_3,\r,t)]\biggl\},\label{dissipative:R}
\ee
where $\p_c=\hbar\k_c$ and $\epsilon_c=\hbar\omega_c$ is  a condensate momentum and energy, $f$ is the nonequilibrium distribution function of the noncondensate, and $\hbar\tilde\omega_i\equiv \hbar^2k_i^2/2m+2\tilde g(n_c(\r,t)+\tn(\r,t))$ is a noncondensate energy. The condensate chemical potential and energy are defined by
\be
\mu_c(\r,t)&=&\tilde gn_c(\r,t)+2\tilde g\tn(\r,t),\\
\epsilon_c(\r,t)&=&\mu_c(\r,t)+\frac{m}{2}v_x^2(\r,t)+\frac{m}{2}v_y^2(\r,t)+\frac{m}{2}
\left(\frac{m}{m_{\mu}^*}\right)v_z^2(\r,t).
\ee
In deriving above hydrodynamic equations~(\ref{derivation_dissipative:Josephson_eq}) and (\ref{derivation_dissipative:continuity_eq}), we have expanded the gradient term of the phase regarding the term as small, which is the case in the study of small amplitude oscillations, and introduced the effective mass and the chemical potential effective mass, $m^*\equiv\hbar^2/d^2\left[J-\tilde J n_c(\r,t)\right]^{-1}$ and $m^*_{\mu}\equiv\hbar^2/d^2\left[J-2\tilde J n_c(\r,t)\right]^{-1}$, respectively.
We neglected the $n_c$ dependence of $m^*_{\mu}$, quantum pressure term, and $\nabla R(\r,t)$, in deriving Eqs.~(\ref{derivation_dissipative:Josephson_eq}) and (\ref{derivation_dissipative:continuity_eq}). 
The coupled set of hydrodynamic equations~(\ref{derivation_dissipative:Josephson_eq}) and (\ref{derivation_dissipative:continuity_eq}) are useful starting point for discussing the collective oscillations of the condensate in an optical lattice at finite temperatures.~\cite{ferlaino2002,fertig2005,sarlo2005}

\section{Instability due to the Collisional Damping Process}
In this section, we demonstrate the useful application of the hydrodynamic equations~(\ref{derivation_dissipative:Josephson_eq}) and (\ref{derivation_dissipative:continuity_eq}) derived in Sec.~\ref{sec:generalized_hyd} by investigating the stability of the superfluidity.
We consider the system in which the condensates are moving with a constant velocity~\cite{sarlo2005} and neglect the confining trapping potential.

Nonequilibrium distribution functions in Eq.~(\ref{dissipative:R}) should be determined by solving a kind of kinetic equation. Instead of solving a kinetic equation, we approximate the nonequilibrium distribution function by the static Bose distribution function $f(\p,\r)=\frac{1}{e^{\beta(\tilde{\epsilon}_p-\tilde{\mu})}-1}$,~\cite{williams}
where $\tilde\epsilon_p\equiv\hbar\tilde{\omega}_p\equiv\frac{p^2}{2m}+2\tilde g(n_c+\tn)$ and $\tilde\mu$ is the chemical potential of the noncondensate atoms.
Expanding $\epsilon_c(\r,t)$ around equilibrium, one can linearize $\Gamma_{12}(\r,t)$ as
\be
\delta\Gamma_{12}(\r,t)&=&\frac{\pi \tilde g^2}{2\hbar}\int\frac{d\k_1}{(2\pi)^3}\frac{d\k_2}{(2\pi)^3}\frac{d\k_3}{(2\pi)^3}\delta(\omega_{c,0}+\omega_1-\omega_2-\omega_3)\nonumber\\&&{}\times
\delta(\k_{c,0}+\k_1-\k_2-\k_3)
[1+f(\k_1,\r)]f(\k_2,\r)f(\k_3,\r)\nonumber\\
&&{}\times\beta\left[\tilde g\delta n_c+m\v_{\bot,0}\cdot\delta\v_{\bot}
+m\left(\frac{m}{m_{\mu}^*}\right)v_{z,0}\delta v_z\right]\nonumber\\
&\equiv&\frac{1}{\tau'(\r,t)}\left[\tilde g\delta n_c+m\v_{\bot,0}\cdot\delta\v_{\bot}
+m\left(\frac{m}{m_{\mu}^*}\right)v_{z,0}\delta v_z\right],\nonumber\\
\ee
where we define the relaxation time $\tau'(\r,t)$ by
\be
\frac{1}{\tau'(\r,t)}&\equiv&\frac{\pi \tilde g^2\beta}{2\hbar}\int\frac{d\k_1}{(2\pi)^3}\frac{d\k_2}{(2\pi)^3}\frac{d\k_3}{(2\pi)^3}\nonumber\\
&&{}\times\delta(\omega_{c,0}+\omega_1-\omega_2-\omega_3)
\delta(\k_{c,0}+\k_1-\k_2-\k_3).
\ee
Finally we obtain the linearized hydrodynamic equations
\be
&&\frac{\partial \delta n_c(\r,t)}{\partial t}
+\nabla_{\bot}\cdot\left[n_{c,0}(\r)\delta{\bf v}_{c}(\r,t)\right]
+\frac{\partial}{\partial z}\left[\frac{m}{m^*}n_{c,0}(\r)\delta v_{z}(\r,t)\right]\nonumber\\
&&{}+\nabla_{\bot}\cdot\left[\delta n_c(\r,t){\bf v}_{c,0}(\r)\right]
+\frac{\partial}{\partial z}\left[\frac{m}{m^*_{\mu}}\delta n_c(\r,t)v_{z,0}(\r)\right]
=-\delta\Gamma_{12}(\r,t),\label{collisional_linearized_continuity_eq}\nonumber\\
\\
&&m\frac{\partial}{\partial t}\delta\v_c(\r,t)=-\nabla
\biggl[m\v_{\bot,0}(\r)\cdot\delta\v_{\bot}(\r,t)+m\left(\frac{m}{m_{\mu}^*}\right)v_{z,0}(\r)\delta v_z(\r,t)\nonumber\\
&&{}\qquad\qquad\qquad\qquad\quad+\tilde g\delta n_c(\r,t)\biggl].\label{collisional_linearized_josephson_eq}
\ee
Looking for a plane-wave solution such as ${\rm exp}[i({\bf k}\cdot\r-\omega t)]$ with $\k=(0,0,k)$, we obtain $\omega
\equiv\v_{c,0}^{\mu}\cdot\k+c^*k\equiv\Omega$
in the limit $1/\tau'\to0$, where $c^*\equiv\sqrt{\tilde g n_{c,0}/m(m/m^*)}$ is the sound velocity modified due to the lattice potential and $v_{c,0}^{\mu}\equiv (m/m_{\mu}^*)v_{c,0}$ is the chemical potential group velocity.~\cite{kramer2003,menotti2003}
When we include $1/\tau'$, the dispersion relation is modified as $\omega=\Omega-i\Gamma$. To first order in $1/\tau'$, we obtain the damping term
\be
\Gamma=\frac{1}{2\tau'}\left(1+\frac{m}{m_{\mu}^*}\frac{v_{c,0}}{c^*}\hat\v_{c,0}\cdot\hat\k\right),\label{collisional:damping_term}
\ee
where $\hat\v_{c,0}$ and $\hat\k$ are normal vectors for $\v_{c,0}$ and $\k$, respectively.
From the expression (\ref{collisional:damping_term}) for $\Gamma$, it can be seen that the instability occurs for the mode with $\hat\v_{c,0}\cdot\hat\k=-1$ when $v_{c,0}^{\mu}>c^*$,~\cite{machholm2003,kramer2003,menotti2003} since the damping term $\Gamma$ changes its sign.

\section{Summary}
In summary, we have investigated the stability of the superfluidity of the condensates in an optical lattice by using the generalized hydrodynamic equations. We derived the Landau criterion microscopically and showed that the criterion is modified due to the lattice potential as expected. 
In the present paper, while we only consider the collisional damping process as a destabilizing process, one can also derive the Landau instability due to the Landau damping process, which is also important in the collisionless regime of interest, from the generalized hydrodynamic equations.~\cite{konabe2006_3} 
The generalized hydrodynamic equations derived in the present paper can be used for analyzing the dipole oscillations and its damping in the presence of the noncondensate atoms such as observed in the experiment by Florence group.~\cite{ferlaino2002}

\section*{ACKNOWLEDGMENTS}
S. K. is supported by JSPS (Japan Society for the Promotion of Science) Research Fellowship for Young Scientists.

\end{document}